\documentclass[%
 preprint,
 superscriptaddress,
 amsmath,amssymb,
 aps
]{revtex4-2}
\usepackage{graphicx}
\usepackage{dcolumn}
\usepackage{bm}
\usepackage{csquotes}
\usepackage[english]{babel}
\usepackage{siunitx}
\usepackage{booktabs}
\usepackage{tabularx}
\usepackage{adjustbox}
\usepackage{xcolor}
\usepackage{lineno}

\usepackage{braket}
\usepackage[normalem]{ulem}
\usepackage[caption=false, justification=centerlast]{subfig}
\usepackage{todonotes}

\newcommand{\rom}[1]{\uppercase\expandafter{\romannumeral #1\relax}}

\usepackage{lineno}

\usepackage[%
colorlinks=true,
urlcolor=blue,
linkcolor=blue,
citecolor=blue
]{hyperref}

\usepackage{tabularx}

\newcommand{\UOL}{Department of Physics, The University of Liverpool, Liverpool, L69 3BX, United Kingdom}
\newcommand{\CI}{Cockcroft Institute, Warrington WA4 4AD, United Kingdom}

\newcommand{\UOM}{Department of Physics and Astronomy, The University of Manchester, Manchester M13 9PL, United Kingdom}

\newcommand{\ICMUV}{ICMUV, Universidad de Valencia, 46071 Valencia, Spain}

\newcommand{\FUHC}{Federal University of Health Sciences of Porto Alegre, Porto Alegre, RS, 90050-170, Brazil}

\newcommand{\PKU}{Center for Applied Physics and Technology, HEDPS, and SKLNPT, School of Physics, Peking University, Beijing 100871, China}

\begin{document}
	      

\title{100s TeV/m-Level Particle Accelerators Driven by High-density Electron Beams in Micro Structured Carbon Nanotube Forest Channel}

\author{Bifeng Lei}
\email{bifeng.lei@liverpool.ac.uk}
\affiliation{\UOL}
\affiliation{\CI}
\author{Hao Zhang}
\affiliation{\UOL}
\affiliation{\CI}

\author{Cristian Bonţoiu}
\affiliation{\UOL}
\affiliation{\CI}

\author{Alexandre Bonatto}
\affiliation{\FUHC}

\author{Javier Resta‐López}
\affiliation{\ICMUV}

\author{Guoxing Xia}
\affiliation{\UOM}
\affiliation{\CI}

\author{Bin Qiao}
\affiliation{\PKU}
\author{Carsten Welsch}
\affiliation{\UOL}
\affiliation{\CI}

\date{\today}

\begin{abstract}

Solid-state materials, such as carbon nanotubes (CNTs), have the potential to support ultra-high accelerating fields in the $\si{TV/m}$ range for charged particle acceleration. 
In this study, we explore the feasibility of using nanostructured CNTs forest to develop plasma-based accelerators at the $100s~\si{TeV}$-level, driven by high-density, ultra-relativistic electron beams, using fully three-dimensional particle-in-cell simulations. 
Two different acceleration mechanisms are proposed and investigated: the surface plasmon leakage field and the bubble wakefield.
The leakage field, driven by a relatively low-density beam, can achieve an acceleration field up to $\si{TV/m}$, capable of accelerating both electron and positron beams. 
In particular, due to the direct acceleration by the driver beam, the positron acceleration is highly efficient with an average acceleration gradient of $2.3~\si{TeV/m}$. 
In contrast, the bubble wakefield mechanism allows significantly higher acceleration fields, e.g. beyond $400~\si{TV/m}$, with a much higher energy transfer efficiency of $66.7\%$. In principle, electrons can be accelerated to PeV energies over distances of several meters.  
If the beam density is sufficiently high, the CNT target will be completely blown out, where no accelerating field is generated. Its threshold has been estimated. 
Two major challenges in these schemes are recognised and investigated. 
Leveraging the ultra-high energy and charge pumping rate of the driver beam, the nanostructured CNTs also offer significant potential for a wide range of advanced applications.
This work represents a promising avenue for the development of ultra-compact, high-energy particle accelerators. 
We also outline conceptual experiments using currently available facilities, demonstrating that this approach is experimentally accessible.
\end{abstract}

\maketitle


\section{Introduction}

The development of ultra-compact plasma-based particle accelerators is primarily beneficial from the ultra-high acceleration gradient, which is achieved through coherent plasma wave excitation driven by high-intensity beams, such as photon or charged particle beams. 
The acceleration field in plasma wave is dependent upon the plasma density as $E = m_e c \omega_p /e \simeq 9.6 \sqrt{n_e [10^{22} \si{cm^{-3}}]} ~[\si{TV/m}] \propto n_e^{1/2}$ where $n_e$ is the ambient plasma electron density, and $m_e$ and $c$ are rest electron mass and speed of light in vacuum respectively, and $\omega_p = \sqrt{4\pi e^2 n_e/m_e}$ is the plasma frequency~\cite{Esarey2009phy}. 
The current state-of-the-art gaseous plasma-based accelerators, for example, laser-driven wakefield accelerator (LWFA) or beam-driven wakefield accelerator (PWFA), practically work with the low-density classical plasma in the range of $n_e \sim 10^{14-18} \si{cm^{-3}}$. This density can, in principle, support an acceleration gradient of approximately $G \sim 1-100 ~\si{GeV/m}$~\cite{Tajima1979, Chen1985acc, Adli:2018aa, Gonsalves2019pet}. 
To achieve a higher $G$, denser plasma is required, which, for example, naturally exists in metallic crystals. The density of charge carriers (conduction electrons) in these crystals ranges from $10^{20-24} \si{cm^{-3}}$ that can support collective plasma waves excited by external sources, e.g. photons or fast electron beams, referred to as plasmon~\cite{Ritchie1957pl}.  
With a high-intensity beam driver, the plasmon amplitude can be extremely high, in the order of TV/m. Such intense fields could be used for the charged particle acceleration~\cite{Tajima:2014aa, Gilljohann:2023aa, Chattopadhyay:2020aa, Sarma:2022aa}.
However, the solid-density plasma wakefield acceleration using natural crystals remains unrealisable due to the limitation of angstrom-size channels and the unavailability of suitable drivers, e.g. high-intensity x-ray lasers~\cite{Shin:2019aa,Zhang2016par}.  


In recent years, the advent of modern nanofabrication techniques has enabled the fabrication of nanostructures and metamaterials with enhanced flexibility in parameters such as structure, density, and thickness compared to their unstructured crystalline counterparts, achieved through the controlled deposition of porous materials~\cite{Gleiter:2000aa, Xiao:2020aa}.
Recent numerical simulations have demonstrated that pure graphene layers of plasma density $10^{22}~\si{cm^{-3}}$ can achieve a $4.79~\si{TeV/m}$ acceleration gradient for electron acceleration by utilising a high-intensity, $10^{21}~\si{W/cm^{2}}$, ultra-violet laser to drive plasma wakefield~\cite{Bontoiu:2023aa}. Nevertheless, the preparation of such laser pulses is still a challenging process~\cite{Naumova2004re, Mourou:2014aa}.

A carbon nanotube (CNT) is a graphene-based single- or multi-walled tubular structure. With a typical inner diameter of a few nm and a length of up to mm, CNTs exhibit excellent mechanical, thermodynamic and electronic properties~\cite{Eatemadi:2014aa, Rathinavel:2021aa}. 
Depending on their chirality, CNTs can be metals (armchair) which can provide wall plasma densities in the range of $10^{22-24}~ \si{cm^{-3}}$ with a vacuum hollow channel permitting the channelling of optical laser pulses or particle beams over extended distances to excite the strong plasmons on the wall surface, referred as surface plasmons (SPs)~\cite{Stern1960sur,Tajima:1978aa, Ukhtary:2020aa, Martin-Luna:2024ab}.
However, at the current stage, it is also not feasible to directly employ these CNTs for particle acceleration due to the inability to resonantly match the $\si{\mu m}$ dimensions of available photon or particle beams with a few $\si{nm}$ natural dimensions of CNTs. 

\begin{figure}
	\includegraphics[width=0.45\textwidth]{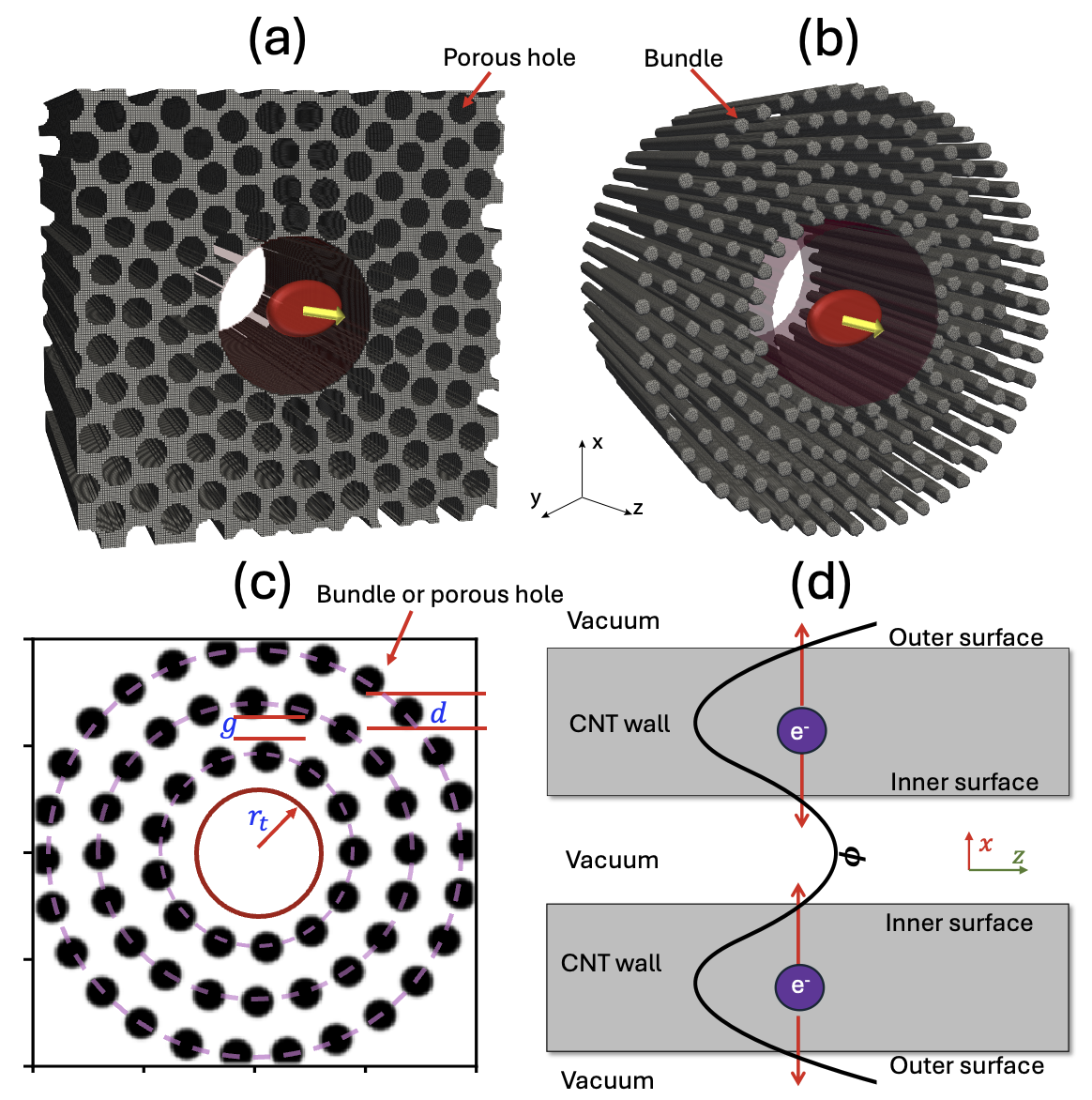}
	\caption{Schematic illustration of (a) porous and (b) bundle CNT targets, arranged in the form of CNT arrays. The boundary of the vacuum channel is indicated by the light red surface surrounding the axis, while the driver beam is represented by the red spots with propagation direction indicated by yellow arrows. Each bundle or the porous wall could be composed of hundreds of single-walled or multi-walled carbon nanotubes.
	(c) Transverse layout of structured CNT target implemented in simulations. The black spots indicate the locations of CNT bundles or pores which occupy each centralised virtual layer by a filling factor $f_f$. The red circle indicates the position of the inner surface.
	(d) A slice of nanostructured CNT target. The conduction electrons (purple spots) have the probability to transit across the inner and outer surfaces.
	The black curve represents the transverse potential profile.}
	\label{fig:cnt_targets}
\end{figure}

Nanostructured CNT targets, e.g. porous or bundled CNT as shown in Fig.~\ref{fig:cnt_targets}, arranged as dense CNT arrays or forests, offer a greater degree of flexibility in geometric structure, density, and surface structures than unstructured solids~\cite{Futaba:2006aa,Iijima:1991aa, Yang:2018aa}.
This is due to the unique characteristics of CNT arrays, which can be tailored more specifically through controlled synthesis and manipulation of their structural and compositional properties.
The wall density being adjusted to a range of $n_t \sim 10^{19 - 24} \si{cm^{-3}}$ can facilitate a $0.1-100~\si{TV/m}$-level accelerating and focusing field. 
The diameter of the vacuum hollow can range from hundreds of nm to a few $\si{\mu m}$, enabling the channelling of either a photon beam (such as an optical laser pulse) or a high-energy charged particle beam for a sufficient length of time to drive solid-state SPs, while simultaneously mitigating adverse effects, e.g. collision~\cite{Chen1987as}, beam filamentation~\cite{Benedetti:2018aa}, bulk radiation~\cite{Zong1995RadiationIE}, etc. 
In modern laboratories, porous CNT targets can be prepared using the so-called anodic aluminium oxide (AAO) template method, which provides excellent tunability through control of the AAO template's pore texture~\cite{Thompson:1997aa, Kyotani:2006aa,Hou:2012aa}.
Moreover, the preparation of CNT bundle targets can be achieved through several methods, e.g. chemical vapour deposition (CVD), arc discharge or template-assisted growth. Each of these methods offers a distinct level of control over the properties of the resulting CNTs and their bundles~\cite{Joselevich:2008aa, See:2007aa, Schifano:2023aa, Schifano:2023aa, Yadav:2024aa}.
Coherent oscillations of the conduction electron gas in these nanostructured CNTs can be excited when an ultra-relativistic, high-density electron beam (referred to as the driver beam) passes through the vacuum channel (see Fig.~\ref{fig:cnt_targets}). During this process, conduction electrons within the CNT walls are radially compressed and gain transverse momentum.
The key distinction between nanostructured CNT-based plasma and the uniform plasma typically used in LWFA and PWFA experiments lies in the inner and outer surfaces of these hollow targets, as depicted in Fig.~\ref{fig:cnt_targets}(d). 
The conduction electrons may be trapped inside the CNT wall by potential barriers (vertical curve) formed across these surfaces, leading to unique electron dynamics, such as excitation of SPs.  
Whether the electrons can cross these surfaces determines the wakefield dynamics.

When the driver beam density is moderate, the compressed electrons may rebound due to the space charge force.  If their inward momentum is sufficient to overcome the potential barrier on the inner surface, the electrons can be emitted into the vacuum region, generating a strong electrostatic field on the order of TV/m, suitable for particle acceleration~\cite{Bonatto:2023aa}.
Here, an individual conducting electron moves freely as being in a uniform plasma and can generate a plasma wave or plasma bubble wakefield in the vacuum channel.
Using an effective density model~\cite{Bonatto:2023aa}, the field amplitude of the bubble wakefield has been theoretically predicted to approach the coherence limit of collective fields as $E_{0}\simeq m_e c \omega_p/e$. This limit, also known as cold nonrelativistic wakefield breaking field~\cite{Esarey2009phy, Dawson:1959aa}, is significantly higher than the amplitude of SP~\cite{Ukhtary:2020aa} and exceeds the lattice ionic field, where polaritons play a negligible role in the acceleration process~\cite{Hakimi2018wa}.
Conversely, if the driver beam density is low and the conduction electrons lack sufficient momentum to overcome the potential barrier, they accumulate periodically on the inner surface, forming conventional SPs. 
In cases where the CNT wall has a finite thickness and the driver beam density is extremely high, conduction electrons may acquire enough outward momentum to escape from the outer surface, leading to their full ejection from the CNT structure. We define this phenomenon as blowout.

In this paper, we present a $100s~\si{TeV/m}$-level plasma accelerator by channelling an ultra-relativistic electron beam through CNT targets such as those shown in Fig.~\ref{fig:cnt_targets}. Two key acceleration mechanisms are demonstrated: leakage field (Sec.~\ref{sec:leakage}) and bubble wakefield (Sec.~\ref{sec:bubble}) .
We demonstrate for the first time that the electromagnetic (EM) fields generated by SPs can penetrate deeply from the inner surface into the vacuum region, referred here to as the leakage field, providing a sub-TV/m-level automatically phase matched accelerating and focusing field for both electron and positron beams.
This mechanism can be driven by a low-density electron beam and is feasible with currently available experimental facilities.
We further demonstrate that with a moderately intense driver beam, the conduction electrons can freely cross the inner surface and form the bubble wakefield inside the vacuum channel. 
The amplitude of this electrostatic field can reach several hundreds $\si{TV/m}$, which is remarkably an order of magnitude higher than the cold nonrelativistic wakefield breaking field $E_0$. This significant increase in field strength indicates that the plasma wave inside the CNT channel is nonlinear and relativistic.
An ultrashort electron beam with nC-level charge can be self-trapped and accelerated to several GeV over distances as short as $100~\si{\mu m}$, achieving an ultra-high acceleration gradient of about $73.0~\si{TeV/m}$ and a high energy transfer efficiency of $67\%$, as shown in our simulations (Seen in Sec.~\ref{sec:bubble}).
The longitudinal acceleration field inside the vacuum channel is radially uniform and can lead to the conservation of the energy spectrum of the witness beam.
Another notable feature is that the transverse focusing field vanishes within the vacuum region but reaches $10s~\si{TV/m}$-level amplitudes within the CNT walls. This ensures that the quality of the accelerated electron beam is maintained and also suggests the potential for accelerating positively charged particles. 
We also show that this CNT-based acceleration is dynamically stable and can be efficiently scaled with $n_b/n_t$, where $n_b$ is the driver beam density.
Additionally, this configuration minimises radiation reaction effects, as most of the accelerated particles remain confined within the vacuum channel.
This makes such systems highly promising for future high-energy applications, with their proven ability to sustain extreme fields.

\section{Geometry of nanostructured CNT targets and numerical simulation}

The schematic geometry of the nanostructured CNT targets, fabricated by arranging CNT wires into specific array patterns such as porous and bundled configurations, is illustrated in Fig.~\ref{fig:cnt_targets} (a) and (b), respectively. 
Fully 3D particle-in-cell (PIC) simulations were performed using the code WarpX~\cite{Fedeli2022pu} to model these structures.
The transverse layout of the targets is shown in Fig.~\ref{fig:cnt_targets}(c), where the distribution patterns of the bundles or pores are initialised with diameter $d$ in the PIC simulations as centralised virtual layers around the central axis. These layers are characterised by a filling factor, $f_f$, which describes the occupation of each layer by bundles or pores, and a gap, $g$, between adjacent layers.
For a CNT bundled target, as shown in Fig.~\ref{fig:cnt_targets}, the black spots represent the CNT bundles with a density of $n_t$, which is made of hundreds of CNT wires, while the remaining areas are vacuum. For a CNT porous target, the black spots represent the vacuum pores, and the surrounding areas are made of CNT wires with the same density, $n_t$. 
For both cases, a vacuum channel with a radius $r_t$ of a few $\si{\mu m}$ radius $r_t$ is created at the centre to guide the driver beam.  
This set of parameters controls the effective plasma density, which will, in turn,  guide the target design in possible future proof-of-principle experiments. For instance, the effective density of a CNT-bundled target can be estimated by 
 \begin{equation}
 	n_{\text{bundle,eff}} = n_t \frac{N f_f}{1+(N-1)(1+g/d)} \mathrm{,}
 \end{equation}
 and for a CNT porous target, it is
 \begin{equation}
 	n_{\text{porous,eff}} = n_t - n_{\text{bundle,eff}}  \mathrm{,}
 \end{equation}
 where $N$ is the number of layers. 
In the simulations, seven layers are configured with $f_f = 0.65$.
 Unless declared, the initial bundle density is set to $n_t=2\times 10^{22} ~\si{cm^{-3}}$ which gives the cold plasma wavelength $\lambda_{pe}=0.23~\si{\mu m}$.
 The diameter of a bundle or porous hole is $d=200~\si{nm}$ with $g=70~\si{nm}$ gap between each virtual layer. The radius of the vacuum core is $r_t=0.75~\si{\mu m}$. 
 As a result, the effective wall density of bundle and porous targets are equal, given by $n_{\text{eff}}=n_{\text{bundle, eff}}=n_{\text{porous, eff}} =1 \times 10^{22}~\si{cm^{-3}}$.
From here on, we use $n_{\text{eff}}$ to denote the effective plasma density of either porous or bundle CNT wall.
Since our simulations show that a porous CNT target supports a more stable acceleration field compared to bundled targets, most of the simulation results in this paper are based on porous CNT targets, unless otherwise specified for cases involving bundled CNT targets. 
The coherent wakefield limit is estimated based on the wall density $n_t$ as $E_{0}(n_t)=m_e c \omega_{pe} /e \simeq 13.5~\si{TV/m}$, or based on the effective density $n_{\text{eff}}$, $E_{0}(n_{\text{eff}}) \simeq 9.6~\si{TV/m}$.
In the simulations, a $3~\si{\mu m}$-long vacuum region is placed before and after the CNT target to initialise the driver beam and diagnose the emerging accelerated witness electrons, respectively. 
The species of the targets are carbon atoms. The ionisation energy levels of the Carbon atoms are modified with the CNT properties, including weak $\pi$-bond and C-C bond ($\sigma$-bond) energies. The thermionic emission of electrons is negligible by assuming room temperature. The field ionisation is implemented using the ADK method~\cite{Ammosov:1986ab, Mulser:2010aa}. 
 The dimension of the moving window is $4.8~\si{\mu m} \times 4.8~\si{\mu m} \times 6.4~\si{\mu m}$ with $384\times 384 \times 512$ cells in $x$, $y$, and $z$ directions respectively which gives $16$ points in each direction inside each bundle or pores transversely. There are 8 macro particles per cell, which should be sufficient to accurately resolve the electron dynamics inside.

The driver electron beam is characterised by a Gaussian profile with root-mean-square (RMS) dimensions of $\sigma_{r0} = 1.0~\si{\mu m}$ and $\sigma_{z0}= 0.4~\si{\mu m}$ in the transverse and longitudinal directions, respectively, and is represented by $10^7$ macro particles in simulations. 
The beam density is calculated as $n_b=Q_b/[(2\pi)^{3/2} e \sigma_{r0}^2 \sigma_{z0}]$ with $Q_b$ the driver beam charge.
The mean initial energy is $1-10~\si{GeV}$, which is significantly higher than the plasmon energy in CNTs, $\hbar \omega_p=5.4~\si{eV}$.
Such bunches can create hundreds of eV potential over atomic scales, which is sufficient to directly ionise the conduction electrons of carbon atoms in the CNT structure. The ejected electrons can gain kinetic energy far exceeding the Fermi energy~\cite{Pines1953ac, Suzuki2000wo,Ago:1999aa}. Furthermore, the mean free time of collision between electrons and the lattice vibrations is of the order of $\si{ps}$ level, which is much longer than the plasma collective oscillation period at the fs-level.  Therefore, quantum correction for plasma dispersion, the collision effect and the damping effect on the collective oscillation of the conduction electrons can be neglected~\cite{Ritchie1957pl}.

\section{Surface plasmon excitation and leakage field acceleration} \label{sec:leakage}

\begin{figure*}
	\includegraphics[width=0.9\textwidth]{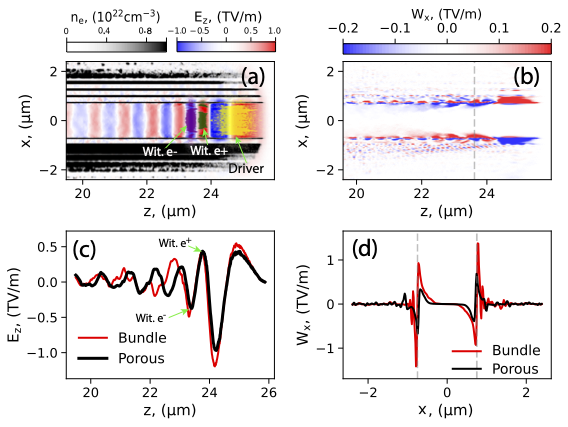}
	\caption{PIC results: Leakage field excitation in nanostructured CNT target. 
	(a) Slice of conduction electron density distribution (grey colourmap), longitudinal electric field $E_z$ (red-blue colourmap), driver and witness electron and positron beams (indicated by yellow, green and purple colours and arrows, respectively). 
	(b) Slice of transverse field $W_x$ along $y=0~\si{\mu m}$. 
	(c) On-axis ($x=0$) plot of $E_z$ for porous (black) and bundle (red) targets.
	(d) Line plot of $W_x$ for porous (black) and bundle (red) targets at $z=23.5~\si{\mu m}$ indicated by the vertical dashed grey line in (b). The two vertical lines indicate the position of the interface of the CNT wall and the vacuum.}
	\label{fig:leakage_field}
\end{figure*}

First, we use a $1~\si{GeV}$ driver electron beam with a total charge of $Q_b = 0.3~\si{nC}$, yielding a beam density of $n_b = 3 \times 10^{20}\si{cm^{-3}}$, or $n_b/n_t = 0.015$. 
This beam is unable to drive the conduction electrons crossing the inner surface, as shown in Fig.~\ref{fig:leakage_field}(a), where plasma oscillations remain confined to the surface.
However, the electromagnetic field of the SPs can leak into the vacuum region, as shown in Fig.~\ref{fig:leakage_field}(a) and (b). The longitudinal leakage field forms a travelling wave that follows the driver beam, with an amplitude up to $1.0~\si{TV/m}$, as seen in Fig.~\ref{fig:leakage_field} (a) and (c). 
The wavelength is $\lambda_{\text{SP,PIC}} = 0.52~\si{\mu m}$. This is in good agreement with the theoretical resonant SP wavelength, $\sqrt{2} \lambda_{\text{eff}} \simeq 0.47~\si{\mu m}$~\cite{Ritchie1957pl}, where $\lambda_{\text{eff}}$ is calculated based on the effective wall density $n_{\text{eff}}$.
Therefore, this field is suitable for sub-$\si{fs}$-level charged particle acceleration.
The acceleration field within the channel is observed to be radially uniform, ensuring that the energy spectrum and beam divergence of the witness beam remain preserved throughout the acceleration process.


The transverse field, e.g. defined as $W_x = E_x - cB_y$ in $x$-direction, where $E_x$ and $B_y$ are the transverse electric and magnetic fields in $x-$ and $y-$directions, respectively, can also shallowly leak into the vacuum region, providing a confining force for both positrons in the vacuum and also preventing electrons from crossing the inner surface, as shown in Fig.~\ref{fig:leakage_field}(d). 
The case of the transverse field in the $y$-direction is similar and will not be shown here.
If an electron lacks sufficient transverse momentum, it will be confined either within the CNT wall or in the vacuum region. 
The transverse field is negligible in the centre region of the vacuum channel, which is highly beneficial as it enables the acceleration of both positively and negatively charged beams while preserving beam quality.
Notably, the bundle CNT target also supports a similar leakage field structure, with marginally higher field amplitudes than the porous target. However, it exhibits unstable oscillations, as seen in Fig.~\ref{fig:leakage_field}(c) and (d). This variation suggests that the internal structure of the CNT wall influences the properties of the leakage field and should be optimised.

\begin{figure}
	\includegraphics[width=0.8\textwidth]{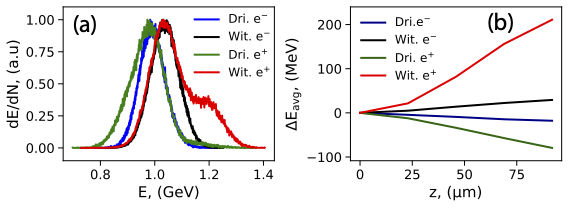}
	\caption{Two PIC results of electron acceleration in (a)-(b) and positron acceleration in (c)-(d). In both simulations, the drivers are electron beams (Blue lines) with the same initial configuration. The witness (Black lines) is electron and positron beams, respectively.  (a) and (c) present the energy spectrum after $89~\si{\mu m}$ propagation in CNT channel. (b) and (d) is the evolution of the average energy, respectively.}
		\label{fig:leakage_energy}
\end{figure}

To test the acceleration process for both electrons and positrons, two simulations are conducted: one with a witness electron beam following the driver beam and the other with a witness positron beam. The witness beams have identical parameters to the driver beam, except for their reduced charge of $0.01~\si{nC}$ and shortened length of $0.1~\si{\mu m}$. 
Both witness beams are initially placed in the first acceleration phase, positioned at $1.44~\si{\mu m}$ behind the driver beam for the electron case and $1.18~\si{\mu m}$ for the positron case, as shown in Fig.~\ref{fig:leakage_field}(a) and (c). 
The energy spectrum of the witness electron and the driver beams are compared in Fig.~\ref{fig:leakage_energy}(a). The witness beam gained energy, while the driver beam lost energy. The mean energy of the driver beam decreased by $16.7~\si{MeV}$, yielding an average energy depletion rate of $R_D = 0.19~\si{TeV/m}$. 
The high energy depletion is primarily driven by excitation of the bulk and surface plasmons and determined by the beam-to-target density ratio $n_b/n_t$~\cite{Ritchie1957pl}.
Simultaneously, the mean energy of the witness electron beam is increased by $29.7~\si{MeV}$, as shown in Fig.~\ref{fig:leakage_energy}(b), leading to a mean acceleration gradient of $G = 0.33~\si{TeV/m}$.
For the witness positron beam as shown in Fig.~\ref{fig:leakage_energy}(c) and (d), there is an additional energy gain directly from the driver beam, which leads to an additional energy gain, and the mean energy is increased to $210~\si{MeV}$.  The mean acceleration gradient can reach up to $G =2.36~\si{TeV/m}$. In this case, the corresponding average energy depletion rate of the driver beam is $R_D = 1.12~\si{TeV/m}$.
These results align well with the leakage field amplitudes shown in Fig.~\ref{fig:leakage_field}(a) and (c), demonstrating field strengths 1–3 orders of magnitude higher than the acceleration gradients typically achieved in gaseous plasma-based LWFA and PWFA\cite{Gonsalves2019pet, Adli:2018aa}.
Moreover, the relatively low $n_b/n_t$ suggests that such an experiment is feasible with existing facilities. For example, with a CNT target density of $n_t = 10^{20},\si{cm^{-3}}$, a beam density as low as $n_b = 10^{18},\si{cm^{-3}}$ would be sufficient conditions achievable at facilities like FACET-II at SLAC~\cite{Yakimenko2019facet} or in the upgraded CLARA facility at Daresbury Laboratory~\cite{Snedden2024spe}.
It should be noted that a condition for acceleration in the SP leakage field is that the witness electron must be relativistic. This enables flexibility in using external injection.

However, it is important to also note that the acceleration gradient $G$ is still 2 orders of magnitude lower than the coherent field limit $E_{0}$ predicted for the CNT wall density $n_t$ used in these simulations. This indicates that only a small fraction of the plasmon energy is leaking into the vacuum channel for acceleration, resulting in relatively low energy efficiency.
The energy transfer efficiency, defined as the ratio of energy gained by the witness beam to the total energy lost from the driver beam after $89~\si{\mu m}$ propagation, is given for positron acceleration by
 \begin{equation}
	\eta_{SPs} = \frac{\int_{V_b} \mathcal{E}_{\text{b}} n_b d \bm{r}}{\int_{V_w} \mathcal{E}_{\text{w}} n_w d \bm{r}} \simeq 7.1\%
	\mathcal{,}
	\label{eq:transfer_efficiency}
\end{equation}
where the symbols $b$ and $w$ denote driver and witness beam respectively. $\mathcal{E}_{b,w}$ denotes the energy spectrum, $n_{b,w}$ the beam density distribution and $V_{b,w}$ the beam volume, respectively.
For the electron case, $\eta_{SPs}=6\%$.
The relatively low strength of the acceleration field and the transfer efficiency are attributed to the fact that the conduction electrons, as the primary energy carrier, cannot transit across the inner surface.

\section{Bubble wakefield excitation and acceleration} \label{sec:bubble}

\begin{figure*}
\centering
	\includegraphics[width=0.9\textwidth]{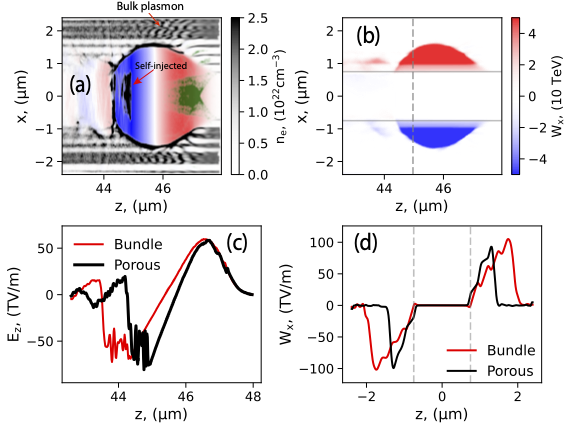}
	\caption{PIC results: Bubble wakefield excitation in nanostructured CNT targets. (a) Slice of conduction electron density distribution (grey colourmap), $E_z$ (red-blue colourmap), and driver beams (green colourmap).(b) Slice of $W_x$. (c) On-axis ($x = 0$) plot of $E_z$ for porous (black) and bundle (red) targets. (d) Line plot of $W_x$ for porous (black) and bundle (red) targets at $z = 48~\si{\mu m}$, indicated by the vertical dashed grey line in (b).}
	\label{fig:wakefield}
\end{figure*}
To drive the conduction electron crossing the inner surfaces, a higher-density driver beam with a total charge of $10~\si{nC}$ and a density of $1.0\times 10^{22}~\si{cm^{-3}}$ or $n_b/n_t=0.5$ is considered. The initial beam energy is $10~\si{GeV}$.

The nonlinear plasma wakefield, also known as bubble wakefield, is excited inside the vacuum channel after the driver beam travels $47~\si{\mu m}$ through the CNT target as shown in Fig.~\ref{fig:wakefield}. This process also triggers self-injection. The resulting plasma bubble is similar to that in gaseous plasma~\cite{Esarey2009phy, Macchi:2020aa, Palastro:2021aa}, with the high amplitude of the acceleration field, reaching up to $83~\si{TV/m}$, exceeding the cold nonrelativistic breaking field limit $E_{0}$.
This indicates that the plasma wave excited in the channel is nonlinear and ultra-relativistic where, in the cold fluid theory, the maximum field amplitude can reach to $E_{WB}=\sqrt{2}(\gamma_p -1 )^{1/2} E_{0}$ where $\gamma_p=(1-v_p^2/c^2)^{-1/2}$ with $v_p$ is the phase velocity of the plasma wave~\cite{Akhiezer1956THEORYOW, Esarey2009phy}.
With the plasma and beam parameters used in the simulations, $E_{WB}$ can reach up to $E_{WB}\simeq 2~\si{PeV/m}$, which is much higher than the numerical value observed in the simulations. This indicates that the cold fluid theory is valid for our situation~\cite{Esarey:1995aa, Schroeder:2005aa}.
The longitudinal acceleration field is also radially uniform, which enhances beam quality.
This amplified field strength results from the confinement of transited wall electrons, preventing them from escaping the vacuum channel, as well as from mitigating bubble breaking.
Unlike typical uniform plasma bubbles, the acceleration volume extends $2~\si{\mu m}$ in the longitudinal and $3~\si{\mu m}$ in the transverse dimensions, which is significantly larger than that of a typical plasma wave of the same density. 
This larger volume supports the acceleration of high-charge witness beams with sizes that fit the aforementioned dimensions.

In the transverse direction, the wakefield does not significantly leak into the vacuum channel but instead creates a strong focusing field within the CNT wall, reaching an amplitude of $100~\si{TV/m}$. 
This phenomenon occurs because the conduction electrons are compressed away from the inner surface, while the transverse wakefield $W_x$ is effectively shielded by the ionic lattice.
As seen from Fig.~\ref{fig:leakage_field}(b)-(c) and Fig.~\ref{fig:blowout}(b)-(c), the transverse field in the vacuum region minimized and the witness beam will undergo very weak transverse oscillation with very low frequency which is proportional to $\sqrt{\nabla \cdot W_x}$. Therefore, the radiation effect should not be relevant in this scenario. 
Even though the peak of the electric field in the channel can reach extremely high values, it is still 4 orders lower than the QED critical field limit. Therefore, the QED effects, such as vacuum polarisation or pair production, are not important for the regimes investigated in this manuscript.
This feature makes the system promising for both electron and positron acceleration within the vacuum region, where the transverse field vanishes~\cite{Schroeder:2013aa, Gessner:2016aa, Yi:2014aa}. It also helps the driver beam channelling stably. 
For positron acceleration, the optimal phase is located in the second plasma bubble around $z = 44~\si{\mu m}$, as shown in Fig.~\ref{fig:wakefield}(a), where the field amplitude reaches up tp $15~\si{TV/m}$. This value is 2-4 orders of magnitude higher than typical field strengths in gaseous plasmas~\cite{Gonsalves2019pet, Adli:2018aa}. However, in the current simulations, the field volume is relatively small and unstable, highlighting the need for optimisation, such as using lower-density targets or a positively charged driver beam.

Bulk plasmon oscillation is also observed within the CNT wall, with a simulated wavelength of $\lambda_{p,\text{PIC}} = 0.22~\si{\mu m}$, closely matching the theoretical value of $\lambda_{p} = 0.235~\si{\mu m}$~\cite{Ritchie1957pl}.
The bundle target exhibits a similar wakefield structure, as shown in Fig.~\ref{fig:wakefield}(b) and (c), with some differences in the spatial dimensions of the plasma bubble.
This consistency validates the effective density model~\cite{Sahai2017exc,Bonatto:2023aa} while also highlighting the influence of the specific nanostructure on the wakefield dynamics.

\begin{figure*}
	\includegraphics[width=0.85\textwidth]{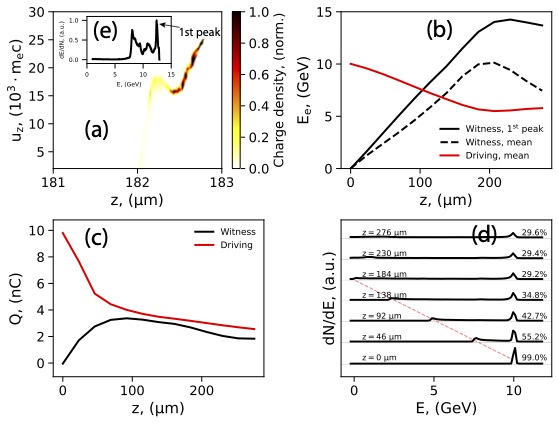}
	\caption{PIC simulation results: (a) Longitudinal phase space of accelerated electrons of the kinetic energy above $200~\si{MeV}$ after leaving the $180~\si{\mu m}$-long porous target.
 	(b) Energy evolution: red solid: mean energy gain of the driver beam; Black solid: peak energy in energy spectrum indicated in (e); black dashed: mean energy of self-trapped electrons of the kinetic energy above $200~\si{MeV}$, respectively.
 	(c) Evolution of charge of the accelerated electrons (black) and driver beam (red).
	(d) Energy spectrum of the driver beam at different propagation distances. 
	(e) Energy spectrum of accelerated electrons.
	The parameters used are the same as those in Fig.~\ref{fig:leakage_field} except for the charge of the driver beam.}
 	\label{fig:wakefield_spectrum}
\end{figure*}


The ultra-high accelerating and focusing field enables the transmitted electrons to be accelerated efficiently and then trapped into the plasma wave by channelling through the vacuum channel for a long distance. 
The energy spectrum of the accelerated electrons, shown in Fig.~\ref{fig:wakefield_spectrum}(a) and (e) after the driver beam propagates $182~\si{\mu m}$, is quasi-monoenergetic, featuring two peaks. The high-energy peak is at $13.0~\si{GeV}$ with a $1\%$ energy spread, and the lower peak is at $8.5~\si{GeV}$ with a $35\%$ spread.
Fig.~\ref{fig:wakefield_spectrum}(b) and (c)  illustrate the energy and charge evolutions of the driver and witness beams, respectively. The energy of the first peak reaches $14.6~\si{GeV}$, corresponding to a peak acceleration gradient of $G=73.0~\si{TeV/m}$. After propagating $200~\si{\mu m}$, the mean energy of the self-trapped electrons attains $10~\si{GeV}$, indicating a mean acceleration gradient of $G=50.0~\si{TeV/m}$. 
Self-trapping saturates around $z = 100~\si{\mu m}$ with a total charge of $3.6~\si{nC}$ of electrons due to significant depletion of the driver beam.
The energy gain is linear before $z=180~\si{\mu m}$ as shown in Fig.~\ref{fig:wakefield_spectrum}(b). This indicates that the dephasing effect is minimised during the acceleration section, and then the energy gain can be linearly scaled until the driver is depleted.

The driver beam rapidly loses energy, with its mean energy dropping to $5~\si{GeV}$, and the charge of electrons above $1~\si{GeV}$ reduces to $3~\si{nC}$. The mean energy transfer efficiency to the witness beam is then estimated as $\eta_{\text{bw}} = 66.7\%$ after a $200~\si{\mu m}$ of propagation.
In this case, the ultra-high energy depletion rate of the driver beam is observed to be $R_D = 25~\si{TeV/m}$, which is $33.3$ times greater than that required for leakage field excitation. 
Additionally, the initial high beam energy contributes to the elevated $R_D$ value, as it extends the lifetime of the strong bubble wakefield excitation. 
The mean charge depletion rate of the driver beam is also significant, reaching $R_Q = 0.35~\si{n C}/ \mu m$.
This highly efficient depletion is shown in Fig.~\ref{fig:wakefield_spectrum}(d), where a significant portion of the driver beam is lost, as indicated by the red dashed line. After $z=200~\si{\mu m}$, only $29\%$ of the high-energy electrons remain insufficient to sustain the strong wakefield. 
At this point, the total charge of the witness beam balances with the driver beam, as shown in Fig.~\ref{fig:wakefield_spectrum}(c). 
This balance also indicates that a higher charge driver beam is required in the wider channel. 
As the energy or charge of the driver beam depletes, the energy transfer ceases, and the witness beam begins to lose energy and charge to the target. 
This indicates the maximum propagation distance of a driver beam inside a CNT channel, referred to as the depletion limit $L_p$, which is crucial for performance. 
To extend $L_p$, higher beam charge $Q_b$ or target density $n_t$ should be considered.

\section{Blowout regime}

\begin{figure}
	\includegraphics[width=0.45\textwidth]{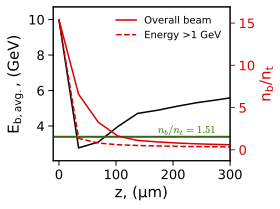}
	\caption{Evolution of mean energy of overall beam (black solid) and evolution of $n_b/n_t$ of overall electron (red solid) and electrons of energy above $1~\si{GeV}$ (red dashed). 
	The horizontal green line indicates the blowout condition.}
	\label{fig:blowout}
\end{figure}

When the driver beam charge $Q_b$, and consequently the beam density, becomes sufficiently high, the wall electrons experience the full blowout, escaping the finite CNT wall structure~\cite{Sahai2017exc}. This occurs when the outward Coulomb force exerted by the driver beam exceeds the restoring force from the wall ions. 
The full blowout conditions can be readily estimated by balancing the beam force as $F_{\text{beam}}\sim n_b V_b >F_{\text{ion}} \sim n_{\text{eff}}  V_{wall}$ where
\begin{equation}
	\frac{n_b}{n_{\text{eff}}} = \frac{V_{wall}}{V_b} =  \frac{\Delta w (2r_t+\Delta w)}{\sigma_{r0}^2} \mathcal{,}
	\label{eq:blowout_condition}
\end{equation}
where $\Delta w = (N-1)g+Nd$ is the wall width. $V_b$ and $V_{wall}$ are transverse beam and wall volume respectively. 
Under the parameters used here, the blowout occurs when $n_b>3.02 n_{\text{eff}}=3.0 \times 10^{22} ~\si{cm^{-2}}$ or $n_b/n_t>1.51$. This implies that the blowout does not occur under the previous simulation conditions.
The blowout regime in plasma wakefield acceleration can be precisely controlled by manipulating the properties of the nanostructured target. For instance, using a thick-walled nanostructure at a fixed wall density, or engineering a high-density Carbon Nanotube (CNT) target when the wall thickness is kept constant, allows for effective control over the blowout dynamics.
Be noticed that since Eq.~\eqref{eq:blowout_condition} is derived based on the effective wall density, $n_{\text{eff}}$, it applies to both porous and bundled CNT targets. Therefore, in the following simulations, only porous targets are considered.

To explore this, a new PIC simulation uses a $200~\si{nC}$ driver beam, giving $n_b/n_t=15.4$.
The energy and charge evolutions are shown in Fig.~\ref{fig:blowout}. 
The blowout happens when the beam density significantly surpasses the threshold $n_b/n_t=1.51$ before $z = 35~\si{\mu m}$, causing all conduction electrons to be stripped away from the CNT wall, preventing the generation of an acceleration field.
At the same time, the beam charge of high-energy electrons is rapidly depleted to below the threshold $n_b/n_t < 1.51$, where the blowout stops and the charge depletion slows down, leaving only $20~\si{nC}$. 
The mean charge depletion rate during the blowout section can be estimated as $R_Q=5.14~\si{n C/ \mu m}$. 
Along with charge loss, the beam energy (black solid line) is also rapidly depleted, particularly for the electrons with energies above $1~\si{GeV}$. 
These electrons lose energy, becoming low-energy electrons, and are subsequently expelled from the vacuum channel by streaming backwards as the confining potential of the inner surfaces prevents their transverse escape.
The observed delay between the charge loss of high-energy electrons (red dashed line) and the overall beam electron loss (red solid line) in the moving window indicates a cascading process. 
In this process, energy and charge are predominantly lost from the head of the beam, while the tail remains largely unaffected, similar to the energy depletion in the plasma dump~\cite{Bonatto:2021aa}.
The bubble wakefield is excited between $35~\si{\mu m} < z < 130~\si{\mu m}$ once the beam density falls below the blowout threshold. During this interval, energy and high-energy charge depletion are significantly reduced. However, as low-energy electrons continue to be lost, the mean energy of the remaining beam increases.
After $130~\si{\mu m}$, the $n_b/n_t$  ratio of high-energy electrons falls below 0.01, resulting in the generation of SPs with lower energy and charge depletion rates.
This behaviour is notable as it demonstrates that the solid-density CNT target scheme is dynamically stable and can be effectively scaled by adjusting the  $n_b/n_t$  ratio, thus offering a promising level of practical robustness.

\section{Discussion and summary}

  \begin{figure*}
  \centering
 	\includegraphics[width=0.9\textwidth]{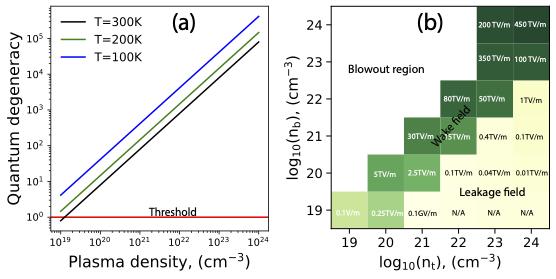}
 	\caption{ PIC results: (a) Diagram of peak acceleration field of porous CNT target with 6 different initial beam $n_b$ and plasma density $n_t$, $1\times 10^{19,20,21,22,23, 24}~\si{cm^{-3}}$. 
 	The yellow colour area indicates the region for the leakage field, the green area for the bubble wakefield, and the white area for the blowout.
 	(b)-(e) Bubble wakefield excitation with four different plasma densities $n_t$ by keeping $n_b/n_t=1.0$. The red-blue colourmap in each plot presents the longitudinal electric field $E_z$.
 	(f)-(h) Longitudinal phase space of accelerated electrons corresponding to three different wall densities $n_t$ where self-injection occurs. 
 	PIC simulations are performed by varying $n_t$ and total beam charge $Q_b$ independently and keeping the other parameters the same. The results are obtained at $z=41~\si{\mu m}$ in the CNT target.
 	}
 	\label{fig:parameter_space}
 \end{figure*}

The presence of the solid surface causes the excitation of SPs to dominate the plasma dynamics. This dominance arises from a balance between the restoring ionic forces of the CNT wall and the Coulomb force exerted by the driver electrons, a balance strongly influenced by the beam-to-plasma density ratio, $n_b/n_t$.
Consequently, the coherent plasma dynamics in the CNT target are significantly different from those in gaseous plasma.
At low beam-to-plasma density ratios (e.g.,  $n_b/n_t < 0.01$), the transverse momentum imparted to wall electrons is insufficient for them to cross into the central vacuum channel. As a result, SP modes are excited along the inner wall surface. In the cylindrical geometry, the longitudinal component of the SP field can leak into the vacuum channel. This provides a sub-$\si{TV/m}$ accelerating field that is automatically phase-matched for both electron and positron acceleration.
When the beam density $n_b$ is moderately high and comparable to the CNT wall density $n_t$, electrons crossing from the wall into the vacuum channel excite a bubble wakefield. This forms structures similar to those observed in gaseous plasmas and generates an electrostatic accelerating field of ultra-high amplitude.
However, exciting this bubble wakefield requires an ultra-high energy and charge pumping rate from the driver beam. This intense pumping can cause energy-depleted driver electrons to stream backwards, which limits the effective acceleration length (maximum pumping length) and degrades the witness beam quality (e.g., causing significant energy spread and charge loss).
The issue of backwards streaming can potentially be mitigated by employing a laser driver.
Additionally, limitations associated with the pumping process can be addressed by optimising the driver beam parameters and target configuration.
Finally, if the beam density exceeds the threshold in Eq.~\eqref{eq:blowout_condition}, the full blowout occurs where no acceleration field exists. While this initially prevents acceleration field formation within the channel, as the driver beam rapidly depletes its energy and its density decreases, it may become possible to sequentially excite bubble wakefields and/or leakage fields, potentially allowing for continued acceleration.
Therefore, achieving high-quality accelerated beams (in terms of charge, energy, and emittance) requires careful initial optimisation of the driver beam and target parameters to ensure stable wakefield generation. Furthermore, well-controlled injection methods are crucial not only for the stability and reproducibility of the acceleration process but also for enabling potential multi-stage acceleration schemes.

As shown in Fig.~\ref{fig:parameter_space} (a), 36 simulations were conducted with 6 distinct values of each $n_t$ and $n_b$, where the white coloured area represents the blowout regime, the green area is the bubble wakefield regime, and the yellow area corresponds to the SPs regime. 
 The corresponding plasma density distribution and the acceleration fields $E_z$ for $n_t=1\times 10^{21}~\si{cm^{-3}}$, $1\times 10^{22}~\si{cm^{-3}}$, $1\times 10^{23}~\si{cm^{-3}}$ and $1\times 10^{24}~\si{cm^{-3}}$ are shown in Fig.~\ref{fig:parameter_space} (b)-(e). With high density, the bubble becomes unstable, and the wavelength decreases. 
It shows that an ultra-high acceleration field can be achieved in the bubble wakefield region, and the field amplitude can exceed $400~\si{TV/m}$ if  $n_b$  and  $n_t$  can be prepared at the $10^{24}~\si{cm^{-3}}$ level. 
 The corresponding energy spectrums of the self-injected electrons in Fig.~\ref{fig:parameter_space} (f)-(h) confirm the capability of these ultra-high acceleration gradients for electron acceleration. They indicate the self-injection threshold for the high density is then given as $n_t>1\times 10^{22}~\si{cm^{-3}}$. In the low-density target, the high beam density is required $n_b>n_t$.
As the excitation of bubble wakefield requires $n_b/n_t \to 1$, and then high-density driver beam is essential. This poses great challenges to the current facilities.

Generating high-quality electron beams via wakefield acceleration often requires precise control over when and where electrons are injected into the accelerating structure. Several methods can potentially achieve this control for self-injection within CNT targets, drawing inspiration from techniques used in gas-based systems. Potential methods include: 
1) Density transitions: creating abrupt changes in the plasma density can trigger electron injection. This provides flexibility in controlling the injection location and timing. Such transitions can be engineered by doping the CNT target with specific atoms at designated positions, physically constricting the channel, or placing a thin film at the channel's entrance.
2) Initial gas section: filling the entrance section of the CNT target with gas enables injection control using methods similar to those established for purely gaseous targets~\cite{Tooley:2017aa}.
3) Channel entrance effects: the sharp physical edge where the driver beam enters the solid CNT channel can itself cause electrons to be injected into the wakefield~\cite{Bontoiu:2023aa}.
4) Laser triggering: an auxiliary laser pulse can be used to precisely time and control the electron injection process~\cite{Kuschel:2018aa}.
5) Ionisation control via doping: introducing specific foreign atoms (dopants) into the CNT material allows control over injection. By selecting dopants that ionise at a specific point (phase) in the wakefield, the injection timing can be managed.
6) External injection: instead of using electrons from the CNT plasma itself, electrons can be supplied by an external accelerator. While not technically self-injection, this method gives excellent control over the initial electron beam's charge, energy, and emittance.
Additionally, the distinct properties of solid-density plasma found in CNTs offer opportunities to explore and develop entirely new injection techniques beyond these established concepts.

This demonstrates the substantial potential for developing CNT-based accelerators capable of accelerating electrons to the TeV energy level. However, achieving such high-density beams poses significant challenges with current state-of-the-art facilities and some critical technological advancements are still required, such as
1) new nano techniques must be developed to grow the target sufficiently long.
2) with the currently available CNT in 10s of $\si{cm}$ long at maximum~\cite{Sugime:2021aa}, strategies to stage hundreds of accelerating sections are required in order to achieve $\si{PeV}$-scale energy. 
3) technologies for producing the required ultra-high-density, high-energy driver beam are not yet fully developed and require further research.
For example, a 1 nC electron bunch with a duration of $100s~\si{attoseconds}$ needs to be a transverse size of hundreds of $\si{nm}$ to achieve a beam density close to $n_b~\sim 10^{24}~\si{cm^{-3}}$.
This level of compression may be possible with future bunch compressor chicanes, such as the one proposed for FACET-II at SLAC~\cite{Yakimenko2019facet}, which anticipates a $125~\si{GeV}$ electron beam with a density up to $10^{34}~\si{cm^{-3}}$.
In the current stage, FACET-II will be capable of delivering  $3~\si{nC}$ electron and positron beams at $10~\si{GeV}$ with a transverse size of down to $0.25-2~\si{\mu m}$ and a duration of $0.5~\si{\mu m}$. This could result in a beam density in the range of $10^{20-22}~\si{cm^{-3}}$, sufficient to drive the bubble wakefields of acceleration gradient on the order of tens of \( \si{TeV/m} \), or even trigger blowout.
With other novel electron sources, such as electron beams from X-ray free-electron lasers facilities~\cite{Huang:2021aa}, or LWFA and PWFA, a relatively weak bubble wakefield can be excited if the beam can be further compressed.
Facilities like CLARA~\cite{Snedden2024spe}, which can provide $250~\si{MeV}$, sub-$\si{nC}$-level electron beams, could also play a role. If such beams can be compressed to $\si{\mu m}$-level by, e.g. plasma lenses~\cite{Pompili2018foc, Hairapetian1994exp} where transverse focusing by 1-2 orders of magnitude is possible, the peak density could reach $10^{18-19}~\si{cm^{-3}}$. This would be sufficient to drive leakage fields with TV/m acceleration, enabling proof-of-principle experiments.
The target length could be extended to tens of micrometres, potentially achieving energy gains of $10s~\si{MeV/\mu m}$.

In summary, one of the key distinctions of the nanostructured CNT target is its solid surface. This contrasts with conventional gaseous hollow plasma channels~\cite{Lindstrom:2018aa, Gessner:2016ab}. This solid boundary enables two unique effects: (1) surface confinement of electromagnetic fields and electrons, and (2) excitation of SPs. As a result, they fundamentally modify the dynamics of beam-plasma interaction in comparison to uniform gas-based systems. It therefore offers CNT-based solid-state plasma accelerators great potential for advancing the development of ultra-compact particle accelerators, opening new avenues for various advanced applications.
For example, the ultra-high acceleration gradient to accelerate both negatively and positively charged particles to several $\si{GeV}$ in $\si{fs}$ time scale. This can enable efficient acceleration of short-lived particles, such as muons, which have a finite mean lifetime of, $2.2~\si{\mu s}$ at rest~\cite{Long:2021aa}.
The ultra-high energy and charge pumping rate make it highly suitable for applications such as high-energy beam dumps or recycle facilities~\cite{Lopez2019des, Bonatto:2021aa}. Combined with excellent mechanical, thermal, and electrical properties of nanostructured CNT targets, this technology offers a highly efficient and safe option for managing high-energy beams.
The ultra-strong EM fields, reaching amplitudes of several hundred \( \si{TV/m} \) over several $\si{\mu m}$ spatial dimensions, highlight its potential for even more advanced applications, such as strong-field physics~\cite{Hattori:2023aa}, astrophysics~\cite{Remington:2005aa}, and particle physics~\cite{Mendonca:2020aa}, where extreme field strengths are crucial for exploring new physical regimes.

However, critical gaps persist in our understanding of key physical processes required to optimise SP excitation for enhanced wakefield strength and stability.
While negative effects like limited depletion length have been explored for homogeneous gaseous plasmas, further investigation is needed to verify driver beam energy loss behaviour in the proposed targets.
Despite advancements in nanofabrication, the CNT-based plasma accelerator presented here necessitates driver beams of extreme quality, posing a significant challenge for practical implementation.
As the wakefield excitation mechanism in our work relies on the collective electron oscillations on the solid inner surface, it imposes a practical requirement on the beam front of the driver to be sufficiently short. 
At the same time, the solid surface also poses a serious challenge to the repetition rate, as the target will be destroyed by the passage of ultrahigh-density ultra-relativistic drivers. This makes a multi-shot operation impractical. While a movable revolver-type target holder may provide a compromise solution, further study is needed to determine the damage threshold of the target after being exposed to an intense electric field.

Emerging beam compression techniques~\cite{Yakimenko2019pro, Pompili2018foc, Hairapetian1994exp} hold promise in delivering high-density particle beams with transverse dimensions on the order of hundreds of nanometers. These advances will significantly contribute to experimental progress, enabling proof-of-concept experiments at currently accessible facilities, such as CLARA and FACET-II.
A practical compromise might involve using doped semiconductor materials, which can provide free plasma electrons with densities around $10^{18}~\si{cm^{-3}}$. This could easily be matched with currently available beams that have spatial sizes of $10~\si{\mu m}$ and similar densities, achieving acceleration gradients in the tens of GeV/m range.

In addition to lasers and negatively charged particle beams, high-energy positively charged beams, such as positrons or protons~\cite{Kimura2011hol,Zhang2016par}—could also potentially drive significant bubble wakefield modes within a CNT channel. International laboratories like SLAC and CERN could provide access to such beams. Given the similar plasma dynamics, it will be important to further investigate whether these beams display distinct differences in beam modulation and wakefield behaviour and if such differences could offer advantages for acceleration and other applications.

\begin{acknowledgements}
Javier Resta-López acknowledges support by the Generalitat Valenciana under grant agreement CIDEGENT/2019/058.
This work made use of the facilities of the N8 Centre of Excellence in Computationally Intensive Research (N8 CIR) provided and funded by the N8 research partnership and EPSRC (Grant No. EP/T022167/1). The Centre is coordinated by the Universities of Durham, Manchester and York.
\end{acknowledgements}

\bibliography{bcnt.bib}

\end{document}